\documentclass[twocolumn]{aastex61}

\submitjournal{The Astrophysical Journal}

\begin{document}

\title{Study of Vertical Magnetic Field in Face-on Galaxies using Faraday tomography}

\author{Shinsuke Ideguchi}
\affiliation{Department of Physics, UNIST, Ulsan 44919, Korea}
\author{Yuichi Tashiro}
\affiliation{University of Kumamoto, 2-39-1, Kurokami, Kumamoto 860-8555, Japan}
\author{Takuya Akahori}
\affiliation{Graduate School of Science and Engineering, Kagoshima University, Kagoshima 890-0065, Japan}
\author{Keitaro Takahashi}
\affiliation{University of Kumamoto, 2-39-1, Kurokami, Kumamoto 860-8555, Japan}
\author{Dongsu Ryu}
\affiliation{Department of Physics, UNIST, Ulsan 44919, Korea}
\affiliation{Korea Astronomy and Space Science Institute, Daejeon 34055, Korea}
\correspondingauthor{Dongsu Ryu}
\email{ryu@sirius.unist.ac.kr}

\begin{abstract}
Faraday tomography allows astronomers to probe the distribution of magnetic field along the line of sight (LOS), but that can be achieved only after Faraday spectrum is interpreted. However, the interpretation is not straightforward, mainly because Faraday spectrum is complicated due to turbulent magnetic field; it ruins the one-to-one relation between the Faraday depth and the physical depth, and appears as many small-scale features in Faraday spectrum. In this paper, employing ``simple toy models'' for the magnetic field, we describe numerically as well as analytically the characteristic properties of Faraday spectrum. We show that Faraday spectrum along ``multiple LOSs'' can be used to extract the global properties of magnetic field. Specifically, considering face-on spiral galaxies and modeling turbulent magnetic field as a random field with single coherence length, we numerically calculate Faraday spectrum along a number of LOSs and its shape-characterizing parameters, that is, the moments. When multiple LOSs cover a region of $\gtrsim (10\ {\rm coherence\ length)^2}$, the shape of Faraday spectrum becomes smooth and the shape-characterizing parameters are well specified. With the Faraday spectrum constructed as a sum of Gaussian functions with different means and variances, we analytically show that the parameters are expressed in terms of the regular and turbulent components of LOS magnetic field and the coherence length. We also consider the turbulent magnetic field modeled with power-law spectrum, and study how the magnetic field is revealed in Faraday spectrum. Our work suggests a way toward obtaining the information of magnetic field from Faraday tomography study.
\end{abstract}


\section{Introduction}
\label{sec1}

Faraday tomography, originally suggested by \cite{bur66} and \cite{bd05}, introduced a revolutionary progress in the study of cosmic magnetic field, superseded by that using Faraday rotation measure (RM). The technique creates a tomographic reconstruction of polarization spectrum as a function of RM or the Faraday depth along the line of sight (LOS). The basic equation is
\begin{equation}\label{eq:PI}
P(\lambda^2)=\int_{-\infty}^{+\infty}F(\phi)e^{2i\phi\lambda^2} d\phi
\end{equation}
where $P(\lambda^2)$ is the observed polarization spectrum. Here, $F(\phi)$ is the Faraday spectrum or Faraday dispersion function (FDF), which is basically the polarized synchrotron emission due to the ``perpendicular'' magnetic field, $B_\perp$, as a function of Faraday depth, $\phi$. The Faraday depth is defined as
\begin{equation}\label{eq:FD}
\phi(x)=K \int_x^0 n_e (x') B_\parallel(x')dx'
\end{equation}
where $B_\parallel$ is the ``parallel'' magnetic field, $n_e$ is the thermal electron density, and $x$ is the physical distance along the LOS. It is given in units of ${\rm rad~m^{-2}}$. The coefficient is $K = e^3/(2\pi m_e^2c^4)$, where $e$ is the electron charge, $m_e$ is the electron mass, and $c$ is the speed of light.

The study of magnetic field using Faraday tomography involves two stages of efforts. One is the reconstruction of $F(\phi)$, and the other is the extraction of magnetic field information from $F(\phi)$. The first requires wide frequency coverage observations of $P(\lambda^2)$, which can be provided by, for instance, the Square Kilometre Array (SKA) and its pathfinders and precursors, such as, LOFAR, ASKAP, MeerKAT, MWA, and HERA. Various approaches for it have been suggested \citep[e.g.,][and references therein]{sun15}. The second requires a successful interpretation of $F(\phi)$. But that often turns out to be difficult, because the physical distance is not in one-to-one correspondence with the Faraday depth owing to the turbulent component of magnetic field, and thus $F(\phi)$, in general, does not represent the distribution of polarized emission in the real space. While it would be straightforward to estimate, for instance, the number of sources of synchrotron radiation and their Faraday depths, Faraday spectrum could be used to obtain more information. The properties of $F(\phi)$ were previously studied \citep[e.g.,][]{bel11,fri11,bec12,ide14b}. For instance, the characteristic features in $F(\phi)$ caused by various configurations of large-scale LOS magnetic field such as field reversal were examined using simple models. Also the effects of small-scale, turbulent field were taken into account and how the effects would be superposed on the features due to large-scale field was studied. It was shown that turbulent field basically appears as many small-scale components in $F(\phi)$, which are called ``Faraday forest'' \citep{bec12}.

In this paper, we extend the second stage efforts. As the first trial, we consider spiral galaxies and study how the properties of ``vertical'' magnetic field (the component vertical to the disk) can be extracted. The strength of vertical magnetic field is among many yet to be constrained in spiral galaxies \citep[see, e.g.,][for a summary]{bec16}. It has been observed in several edge-on galaxies showing the X-shaped pattern \citep[see, e.g.,][]{bec09,kra09}, but such observations so far have told us mostly the existence and orientation of the field. The vertical magnetic field is important for the reconstruction of the global galactic magnetic field and the study of its origin \citep[][]{sof12}, and also necessary to describe the cosmic-ray (CR) transportation (galactic wind). In the Milky Way, there is a difference in the strength of vertical magnetic field toward the north and south Galactic poles, as estimated with RM \citep[][]{tay09,mao10}. This is inconsistent with observations of several external galaxies and needs to be understood \citep[][]{bec16}.

Previously, in \citet{ide14b}, we studied $F(\phi)$ of face-on galaxies, using a realistic model for the Milky Way \citep{aka13}. The model included the global, regular component of magnetic field, based on observations, as well as the turbulent component, constructed with the data of magnetohydrodynamic turbulence simulations. $F(\phi)$ turned out to be complicated, mostly due to turbulent magnetic field; it showed Faraday forest superposed on large scale diffuse emissions, in agreement with \citet{bec12}. We also found that $F(\phi)$ can have significantly different shapes for different configurations of turbulent field, even when the global parameters of the model are fixed. This suggested that while the existence of turbulence can be expected with Faraday forest, it is not easy to quantify the details of turbulence. As a matter of fact, turbulence seems to make it difficult to study the global properties of magnetic field. At the time, our interpretation of $F(\phi)$ was limited because of the complicated behavior. On the other hand, our results indicated that $F(\phi)$ becomes smoother if larger number of LOSs are used.

We, then, attempted to extract the properties of magnetic field using the shape-characterizing parameters of $F(\phi)$, that is, the width, skewness, and kurtosis,
\begin{eqnarray}
\sigma^2 &=& \frac{\sum_l |F(\phi_l)| (\phi_l-\mu)^2}{\sum_l |F(\phi_l)|} \label{eq:sigma} \\
\gamma_{\rm s} &=& \frac{\sum_l |F(\phi_l)| (\phi_l-\mu)^3}{\sigma^3 \sum_l |F(\phi_l)|} \label{eq:skew} \\
\gamma_{\rm k} &=& \frac{\sum_l |F(\phi_l)| (\phi_l-\mu)^4}{\sigma^4 \sum_l |F(\phi_l)|}-3 \label{eq:kurt},
\end{eqnarray}
where $l$ denotes the $l$-th discretized bin of Faraday depth. Here,
\begin{equation}
\mu = \frac{\sum_l |F(\phi_l)| \phi_l}{\sum_l |F(\phi_l)|} \label{eq:ave}
\end{equation}
is the spectrum-weighted average of Faraday depth. We found that stronger vertical magnetic fields result in larger $\sigma$; hence, $\sigma$ should be a useful measure. On the other hand, $\gamma_{\rm s}$ and $\gamma_{\rm k}$ exhibit behaviors too complicated. In summary, in \citet{ide14b}, $F(\phi)$ was obtained with a realistic model, but its behavior was not easy to be interpreted, mainly due to turbulent field.

After the previous work, we here employ ``simple, toy models'' for magnetic field, and try to numerically and analytically describe the behavior of $F(\phi)$. Specifically, for face-on spiral galaxies, we calculate $F(\phi)$ and its shape-characterizing parameters, and examine their dependence on magnetic fields. Even though the models assumed here are simpler than those in former works, they still keep the physical essentials to interpret $F(\phi)$. Most of all, simple models make analytical interpretation possible. We first employ the turbulent magnetic field described as a random field with single coherence length. We also consider the turbulent field represented by power-law spectra. We then examine how $F(\phi)$ obtained along ``multiple LOSs'' can be used to study the vertical magnetic field of face-on spiral galaxies, inspired by the result of \cite{ide14b} that $F(\phi)$ becomes smoother and thus easier to be interpreted with larger number of LOSs. We regard this work as the first step in finding a practical way to extract magnetic field information from Faraday spectrum.

In section \ref{sec2}, we describe our toy model and show calculated $F(\phi)$. In section \ref{sec3}, we present the analytical interpretation of $F(\phi)$. In section \ref{sec4}, we present $F(\phi)$ with power-law, turbulent magnetic field. Summary and discussion follow in section \ref{sec5}. Note that in this paper, we concentrate on the characteristics of ``intrinsic'' $F(\phi)$, and do not consider observational effects in constructing $F(\phi)$ from $P(\lambda^2)$, such as, the ambiguity caused by the limited coverage of observation frequency and observational noises.

\section{Faraday spectrum for random magnetic field with single coherence length}
\label{sec2}

\subsection{Model}
\label{sec2.1}

\floattable
\begin{deluxetable}{cccc}
\tablecaption{Model parameters \label{tab:symbols}}
\tablecolumns{4}
\tablenum{1}
\tablewidth{0pt}
\tablehead{
\colhead{symbol} &
\colhead{physical quantities} &
\colhead{adopted values} &
\colhead{reference}
}
\startdata
$B_{\rm rand}$ & random component of $B_\parallel$ & $\sigma_B = 15/\sqrt{3}\ \mu$G & \cite{bec16} \\
$B_{\rm coh}$ & coherent component of $B_\parallel$ & $0 - 5$ $\mu$G & \cite{bec15,mul17} \\
$n_e$ & thermal electron density & 0.02 ${\rm cm^{-3}}$ & \cite{gae08} \\
$L_{\rm cell}$ & cell size & {\bf 10}, 50, 100 pc & \cite{ohn93,hav06,bec16} \\
$L_{\rm SH}$ & scale height of physical quantities & 0.5, {\bf 1.0}, 2.0 kpc & \cite{gae08,kra09} \\
\enddata
\tablecomments{Fiducial values are denoted with bold.}
\end{deluxetable}

We consider a small portion, $\sim{\rm (100~pc)^2}$, of face-on spiral galaxies, and employ a simple model for galactic magnetic field. The magnetic field is decomposed into the parallel ($B_{\parallel}$) and perpendicular ($B_{\perp}$) components with respect to the LOS. $B_{\parallel}$ contributes to Faraday depth, while $B_{\perp}$ to polarized synchrotron emission, as mentioned in the Introduction. We further assume that the parallel field is decomposed into the random and coherent components, representing the turbulent and global vertical fields, respectively, and express it as,
\begin{equation}
\vec{B}_{\parallel} = \vec{B}_{\rm rand} + \vec{B}_{\rm coh}. \label{eq:Bpara}
\end{equation}

From radio polarimetric data of many {\color{red}almost} face-on external galaxies, the strength of the turbulent magnetic field is estimated to 10 - 15 $\mu$G in spiral arms \citep[see, e.g.,][]{bec16}. So we set the rms (root-mean-square) strength of $B_{\rm rand}$, $\sigma_B$, to be $15/\sqrt{3}\ \mu$G (then, the rms strength of three-dimensional (3D) random field is 15 $\mu$G). It was shown that the size of turbulent cells in the Galactic disk is $\sim$ $10-100$ pc from a pulsar RM study \citep{ohn93}, that the outer scale of turbulent magnetic field is $\sim$ 17 pc in spiral arms and $\sim$ 100 pc in interarm regions from an RM study of extragalactic polarized sources \citep{hav06}, and that the size of turbulent cells in external galaxies is $\sim$ 50 pc from a Faraday depolarization (see below) study \citep{bec16}. Based on these studies, for the coherence length of random field, we adopt 10 pc as the fiducial value and also consider 50 and 100 pc for comparison. On the other hand, we take $B_{\rm coh}$ as a free parameter varying the value, and see the effects on Faraday spectrum. Recent RM studies of face-on external galaxies such as IC 342 \citep{bec15} and NGC 628 \citep{mul17} indicated the absolute values of RM up to $\sim 100\ {\rm rad~m^{-2}}$, which corresponds up to $\sim 6\ \mu$G if we assume the average thermal electron density along the line of sight is 0.02 cm$^{-3}$ and the path length through the thermal gas is 1 kpc. So we set $B_{\rm coh}= 0 - 5\ \mu$G. Note that the positive magnetic field here is meant to be toward the observer, and thus the Faraday depth due to the coherent field is positive.

Regarding the synchrotron radiation, we assume that its polarization angle is the same within the computational domain (see below). This assumption may be justified with observations of ordered magnetic fields in galactic disks \citep[see, e.g.,][]{bec13}. \cite{fle11}, for instance, reported a spiral pattern of ${\vec B}_{\perp}$ from polarized radiation in M51, with an angular resolution of $15''$ which corresponds to the beam size of $\sim$ 500 pc. This means that synchrotron emissions within this beam would have similar polarization angles. From the assumption of the same polarization angle, depolarization caused by the unaligned $B_{\perp}$ within a magnetized, synchrotron polarization emitting medium (wavelength-independent depolarization) does not occur. In addition, depolarization caused by differential Faraday rotation along the LOS within a medium (Faraday depolarization) and that caused by many polarizations with different angles within an observing beam (beam depolarization) are ignored, {\color{red}which means that} ``intrinsic'' Faraday spectrum are considered. This is because the emissions that experience certain Faraday rotation are accumulated in the same Faraday depth. Note that the latter two depolarizations occur in the $P(\lambda^2)$ space.

Thermal and CR electron densities enter in the calculations of Faraday depth and synchrotron emission. Observations suggest $n_e \sim 0.014 - 0.036$ cm$^{-3}$ for the thermal electron density in our Galaxy \citep[see, e.g.,][]{gae08}. So we adopt $n_e =$ 0.02 cm$^{-3}$. We here concern only the overall shape of Faraday spectrum, but not its amplitude (see section \ref{sec2.2}). Hence, we do not need to specify the density and energy spectrum of CR electrons, nor the strength of perpendicular magnetic field, $B_{\perp}$.

The physical quantities described above are assigned to cubic cells of size $L_{\rm cell}$. For $B_{\rm rand}$, we randomly place ${\vec B}_0$ of 15 $\mu$G strength and take the LOS component. Then, $\sigma_B \equiv {\vec B}_0 /\sqrt{3} = 15/\sqrt{3}\ \mu$G, and the cell size corresponds to the coherence length, that is, $L_{\rm cell} =$ 10, 50, 100 pc. Other quantities, such as $B_{\rm coh}$, $n_e$, and the synchrotron emissivity, are assumed to be simply uniform in the computational domain.

Along a LOS, we stack up cells for $[-L_{\rm SH}, L_{\rm SH}]$, where $L_{\rm SH}$ is the scale height of physical quantities. \citet{gae08}, for instance, suggested $L_{\rm SH} \sim 430 - 1830$ pc for the thermal electron density in the thick disk of our Galaxy. \cite{kra09} reported that $L_{\rm SH}$ of radio emission, which reflects $L_{\rm SH}$ of CR electrons and magnetic field, is $\sim 300$ pc for thin disks and $\sim 1.8$ kpc for halos (or thick disks) from observations of various edge-on spiral galaxies. So we set $L_{\rm SH} = 1$ kpc as the fiducial value and also consider 0.5 kpc and 2~kpc for comparison. Each LOS includes $N_\parallel = 2 L_{\rm SH}/ L_{\rm cell}$ cells or ``layers'', that is, 200 layers for representative $L_{\rm cell}$ and $L_{\rm SH}$.

Faraday spectrum is obtained with $N_\perp$ LOSs, covering a small region in the sky where the properties of physical quantities can be assumed to be uniform, but yet larger than the coherence length of turbulent magnetic field. We set each layers to consist of $N_\perp = 1^2 - 30^2$ cells, or the area of $(1^2 - 30^2) L_{\rm cell}^2$. For representative $L_{\rm cell}$, the area becomes $10^2 - 300^2$ pc$^2$, which corresponds to, for instance, $\sim$ $0.''1 - 3''$ for observation of galaxies in the Virgo Cluster at 20 Mpc away.

Our computational domain consists of $N_\parallel \times N_\perp$ cells. With the physical quantities allocated to the cells, we calculate $\phi$ using Equation (\ref{eq:FD}) and the polarized radiation by adding contribution from cells along LOSs, so $F(\phi)$. Below we examine the behavior of $F(\phi)$ for different model parameters, including how the shape-characterizing parameters converge as $N_\perp$ increases.

The model parameters are summarized in Table \ref{tab:symbols}. See section \ref{sec5} for further discussions of our assumptions.

\begin{figure*}
\figurenum{1}
\begin{center}
\vskip -0.8cm
\hskip -1cm
\includegraphics[width=0.58\textwidth,angle=-90]{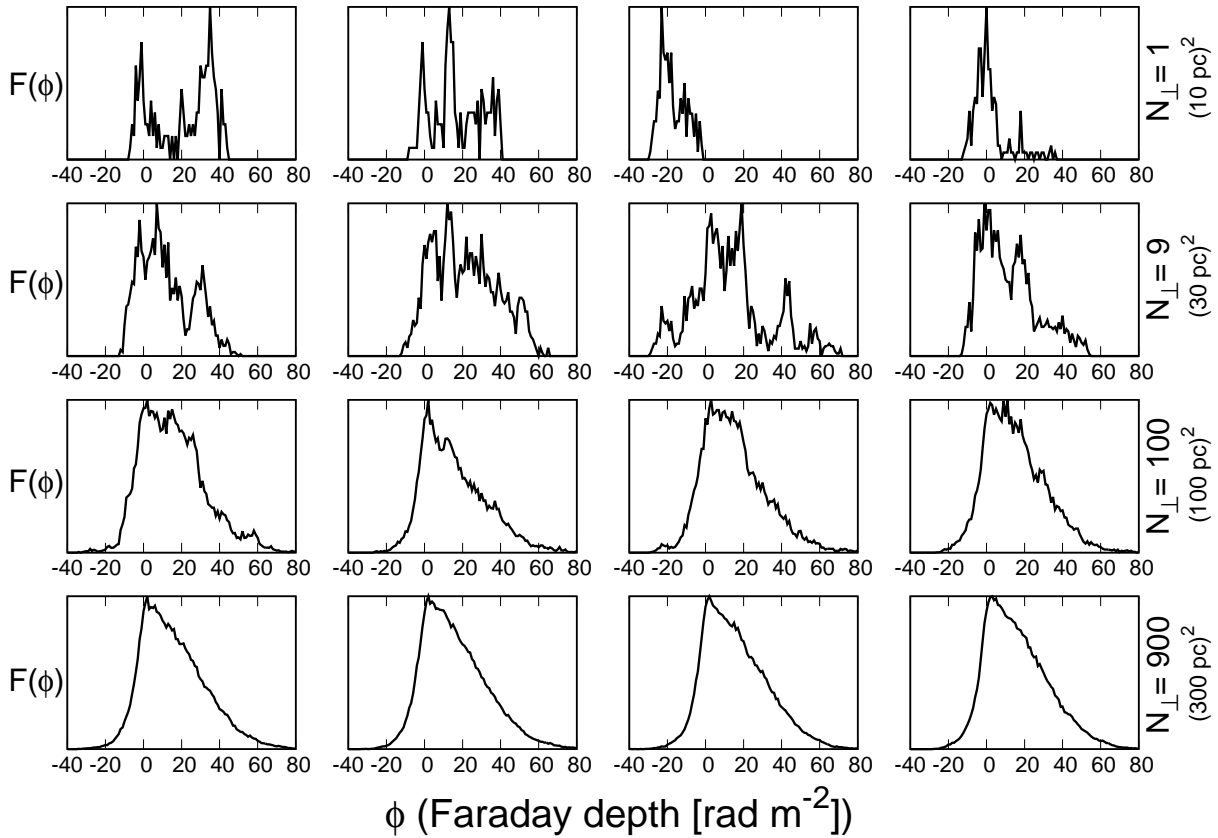}
\end{center}
\vskip 1.3cm
\caption{{\color{red}Simulated Faraday spectrum, $F(\phi)$, as a function of Faraday depth, $\phi$,} with $B_{\rm coh} = 1~{\rm \mu G}$, $L_{\rm cell} = 10$ pc, and $L_{\rm SH} = 1$ kpc, for $N_\perp=$ 1, 9, 100, and 900 from top to bottom. For each value of $N_\perp$, four different realizations are shown.\label{fig01}}
\end{figure*}

\subsection{Results}
\label{sec2.2}

\subsubsection{Convergence}

Figure \ref{fig01} shows $F(\phi)$ with $B_{\rm coh} = 1\ {\rm \mu G}$, $L_{\rm cell} = 10$ pc, and $L_{\rm SH} = 1$ kpc (below representative $L_{\rm cell}$ and $L_{\rm SH}$ are used, unless otherwise stated), for different numbers of LOSs, $N_\perp =$ 1, 9, 100 and 900, from top to bottom. Four different realizations of turbulent magnetic field are shown for each value of $N_\perp$. $F(\phi)$ looks complicated with spikes and varies significantly between different realizations for small $N_\perp$. $F(\phi)$ becomes smooth as $N_\perp$ increases, and converges to a universal shape for $N_\perp \gtrsim 100$. This is because the effects of random magnetic field on $\phi$ are statistically averaged out.

\begin{figure*}
\figurenum{2}
\begin{center}
\vskip -1cm
\hskip -1.6cm
\includegraphics[width=0.52\textwidth,angle=-90]{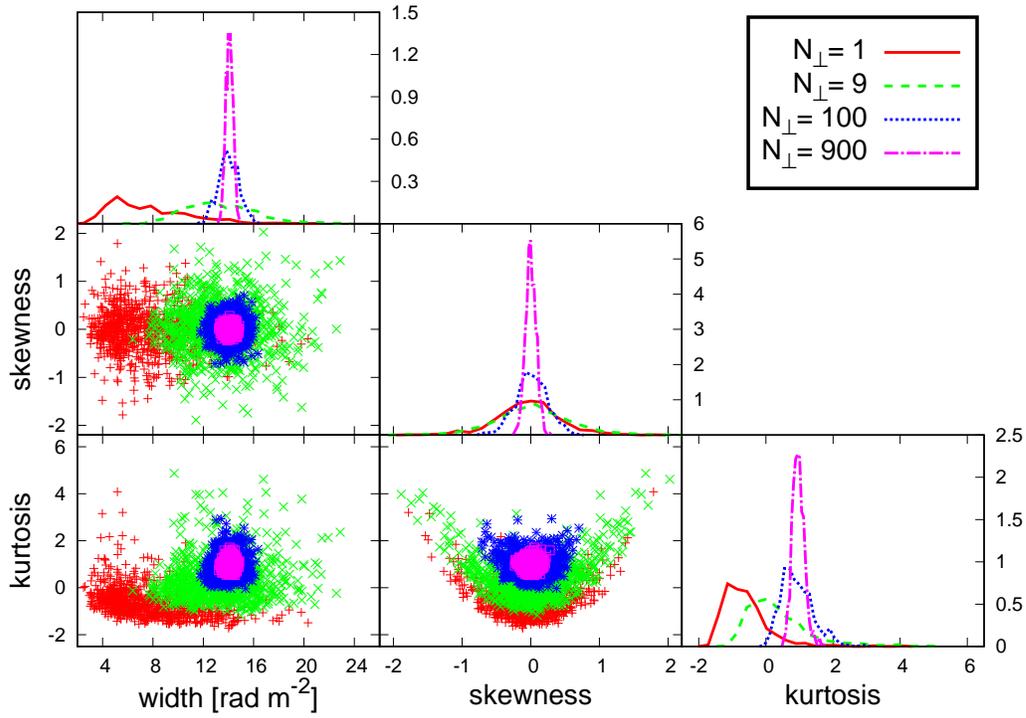}
\end{center}
\vskip 1cm
\caption{Scatter plots for the shape-characterizing parameters of Faraday spectrum with $B_{\rm coh} = 0~{\rm \mu G}$, $L_{\rm cell} = 10$ pc, and $L_{\rm SH} = 1$ kpc, for $N_\perp =$ 1, 9, 100, 900 shown in red, green, blue, and magenta colors, respectively. 800 realizations are shown. Top plots of columns are the one-dimensional probability distributions of the parameters.\label{fig02}}
\end{figure*}

As \citet{ide14b}, the width $\sigma$, skewness $\gamma_{\rm s}$, and kurtosis $\gamma_{\rm k}$ of $F(\phi)$ were calculated (see Equations (\ref{eq:sigma}), (\ref{eq:skew}) and (\ref{eq:kurt})). Figure \ref{fig02} shows the scattered distributions of these shape-characterizing parameters with $B_{\rm coh} = 0~{\rm \mu G}$ for $N_\perp=$1, 9, 100 and 900; 800 realizations for each value of $N_\perp$ are shown. The convergence of the parameters for large $N_\perp$ is evident. The standard deviation of the width, for instance, decreases as $3.09, 2.72, 0.799$ and $0.287$ for $N_\perp=$ 1, 9, 100 and 900, respectively. 

\begin{figure*}
\figurenum{3}
\begin{center}
\vskip -1cm
\hskip -1.8cm
\includegraphics[width=0.52\textwidth,angle=-90]{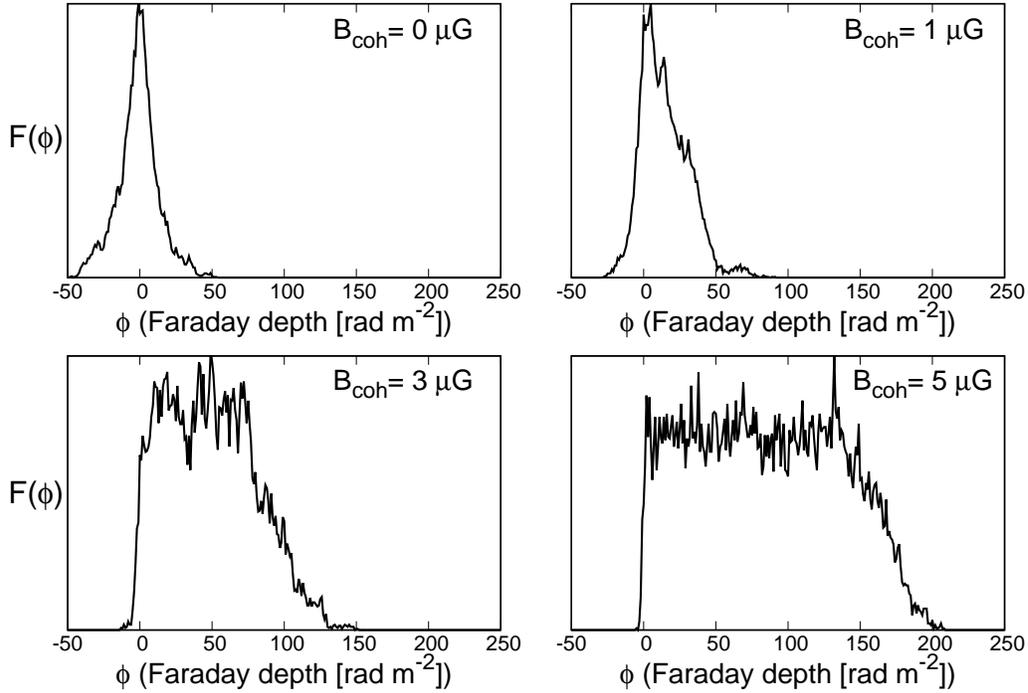}
\end{center}
\vskip 1cm
\caption{{\color{red}Simulated Faraday spectrum, $F(\phi)$, as a function of Faraday depth, $\phi$,} with $B_{\rm coh}=0, 1, 3, 5~\mu{\rm G}$; $L_{\rm cell} = 10$ pc, $L_{\rm SH} = 1$ kpc, and $N_\perp = 100$. 
\label{fig03}}
\end{figure*}

\begin{figure*}
\figurenum{4}
\begin{center}
\vskip -1cm
\hskip -1.6cm
\includegraphics[width=0.52\textwidth,angle=-90]{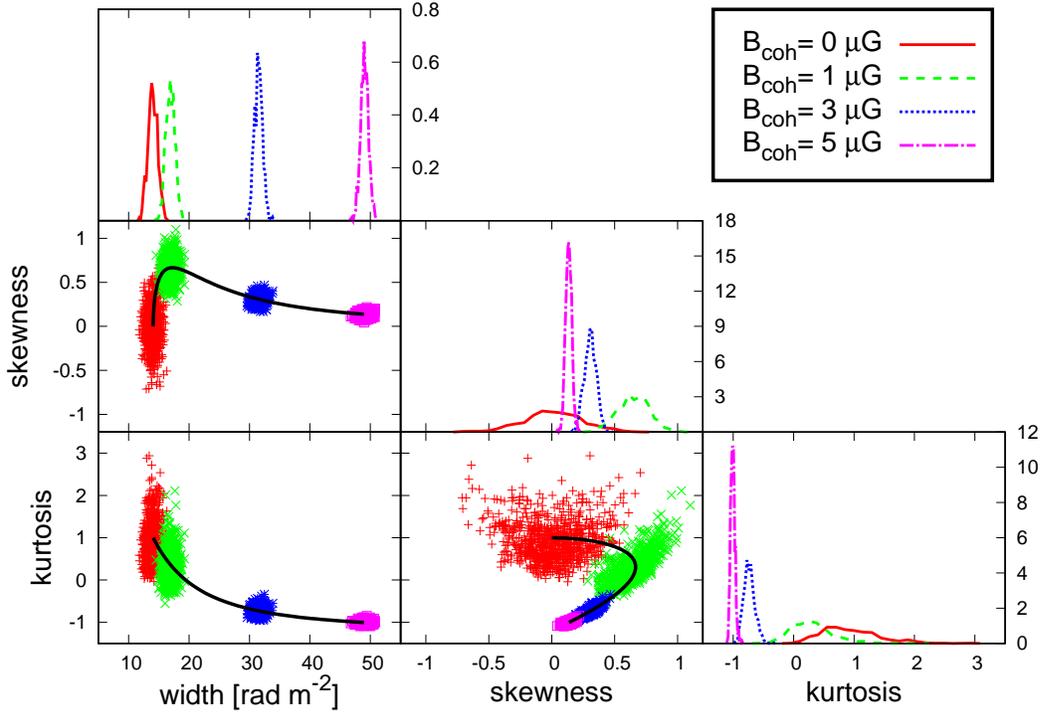}
\end{center}
\vskip 1cm
\caption{Same as Figure \ref{fig02}, but for $B_{\rm coh} = 0, 1, 3, 5~{\rm \mu G}$ shown in red, green, blue, and magenta colors, respectively; $L_{\rm cell} = 10$ pc, $L_{\rm SH} = 1$ kpc, and $N_\perp = 100$. The overlaid black lines show the analytical solutions (see section \ref{sec3}) with $B_{\rm coh}$ varying from 0 to 5 ${\rm \mu G}$. \label{fig04}}
\end{figure*}

\begin{figure*}
\figurenum{5}
\begin{center}
\vskip -1cm
\hskip -1.6cm
\includegraphics[width=0.52\textwidth,angle=-90]{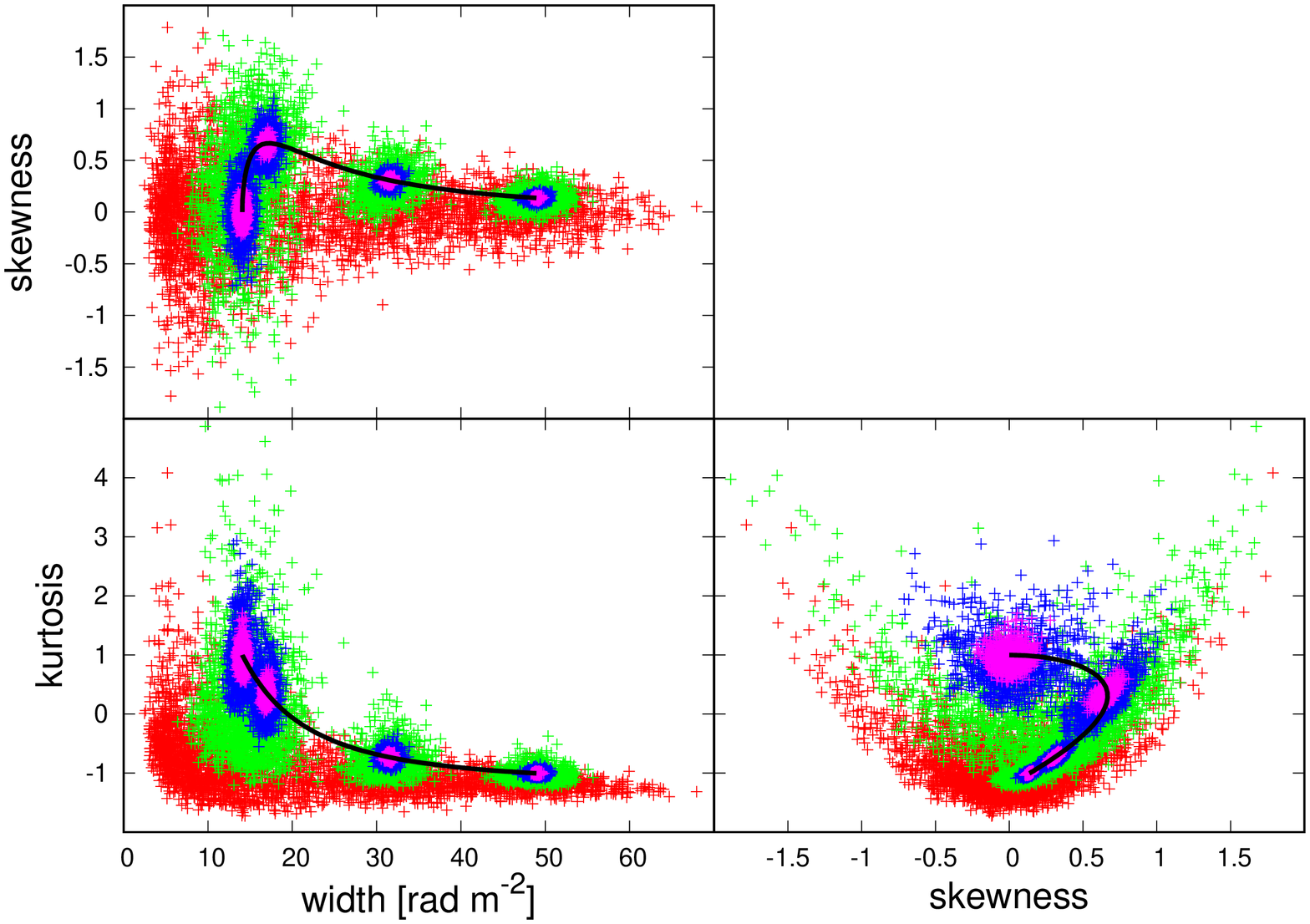}
\end{center}
\vskip 1cm
\caption{Same as Figure \ref{fig04}, but for $N_\perp = 1, 9, 100, 900$ shown in red, green, blue, and magenta colors, respectively. The parameters for different $B_{\rm coh}$, but same $N_\perp$ are plotted with the same color. The overlaid black lines for the analytical solutions are the same as those in Figure \ref{fig04}. One-dimensional distributions are not shown. \label{fig05}}
\end{figure*}

\subsubsection{Dependence on $B_{\rm coh}$}

Figure \ref{fig03} shows $F(\phi)$ for $B_{\rm coh} = 0, 1, 3$, and $5~{\rm \mu G}$, fixing $N_\perp = 100$. The spectrum becomes broader, as $B_{\rm coh}$ increases. With $B_{\rm coh}=0$ $\mu$G, only $B_{\rm rand}$ in layers along LOSs contributes to $\phi$ by the random walk process. On the other hand, with non-zero $B_{\rm coh}$, there is a contribution due to $B_{\rm coh}$ and the contribution monotonically increases along LOSs, on the top of the contribution due to $B_{\rm rand}$. As a consequence, $F(\phi)$ stretches over a large range of $\phi$. The stretching is larger for larger $B_{\rm coh}$. 

Figure \ref{fig04} shows the distributions of the shape-characterizing parameters for $B_{\rm coh} = 0, 1, 3$, and $5~{\rm \mu G}$, fixing $N_\perp = 100$. The parameters change systematically with $B_{\rm coh}$. For $B_{\rm coh}=0$ $\mu$G, $F(\phi)$ has small $\sigma$, as explained. And the narrow, sharp (leptokurtic) shape results in positive $\gamma_{\rm k}$, while the symmetric shape causes $\gamma_{\rm s}$ to be zero. As $B_{\rm coh}$ increases, $\sigma$ increases. At the same time, a flat region appears in $F(\phi)$ (see the $5~{\rm \mu G}$ case in Figure \ref{fig03}), and hence the shape changes from leptokurtic to platykurtic; so $\gamma_{\rm k}$ becomes negative. The change of $\gamma_{\rm s}$, on the other hand, is not monotonic. For non-zero but small $B_{\rm coh}$, $F(\phi)$ becomes positively skewed and $\gamma_{\rm s}$ increases. For larger $B_{\rm coh}$, the flat region restores the symmetry about the mean, causing $\gamma_{\rm s}$ to decrease. These behaviors of $F(\phi)$ and the shape-characterizing parameters will be quantitatively described in section \ref{sec3}. Figure \ref{fig05} shows the convergence of the shape-characterizing parameters for increasing $N_\perp$. The parameters are reasonably converged, again for $N_\perp \gtrsim 100$. This indicates that the observation covering a region of $\gtrsim 100$ times of the square of the coherence length of turbulent magnetic field would be useful to extract the information of magnetic field.

\subsubsection{Dependence on $L_{\rm SH}$ and $L_{\rm cell}$}

\begin{figure*}
\figurenum{6}
\begin{center}
\vskip -0.4cm
\hskip -0.7cm
\includegraphics[width=0.34\textwidth,angle=-90]{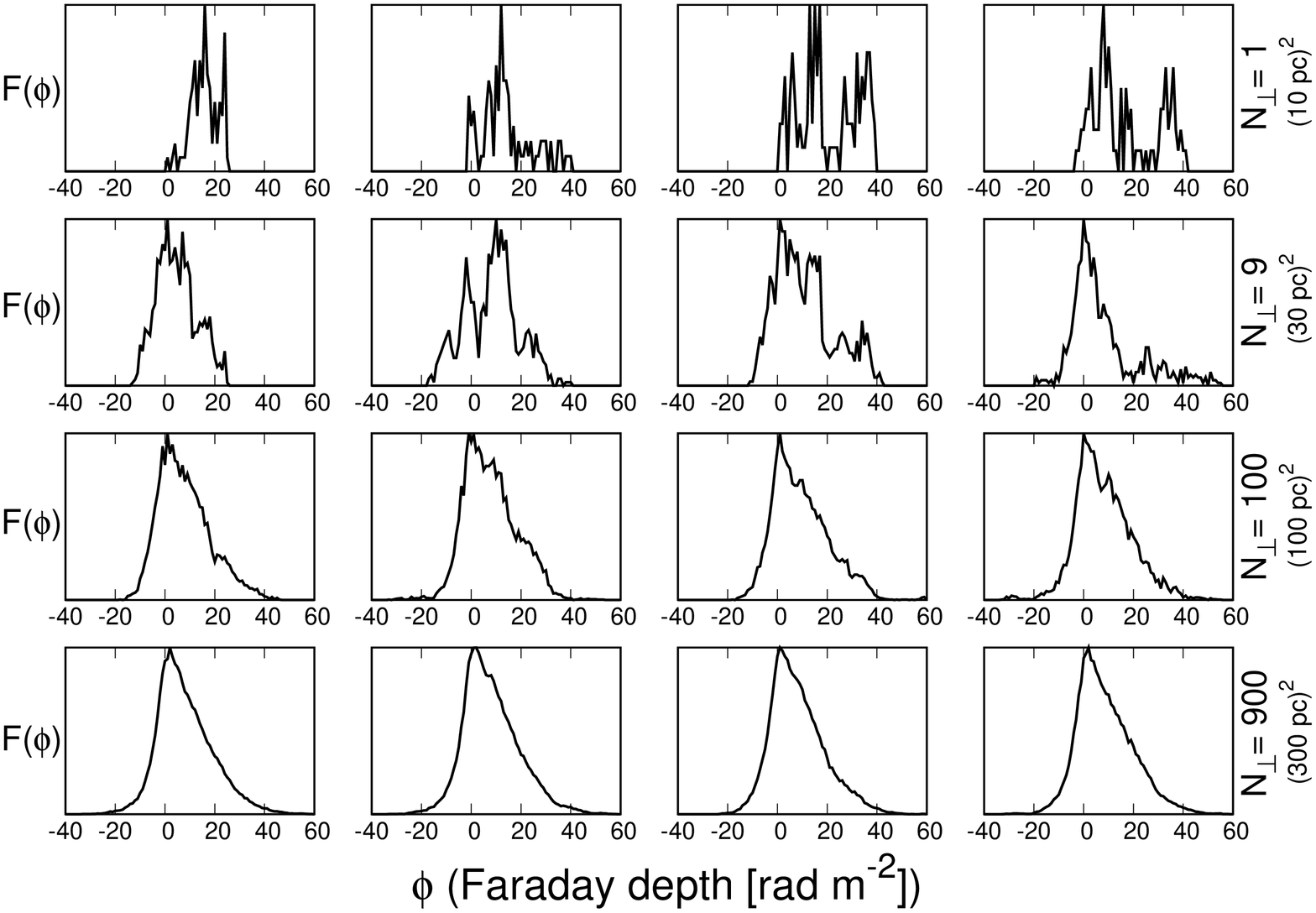}
\hskip 0.6cm
\includegraphics[width=0.34\textwidth,angle=-90]{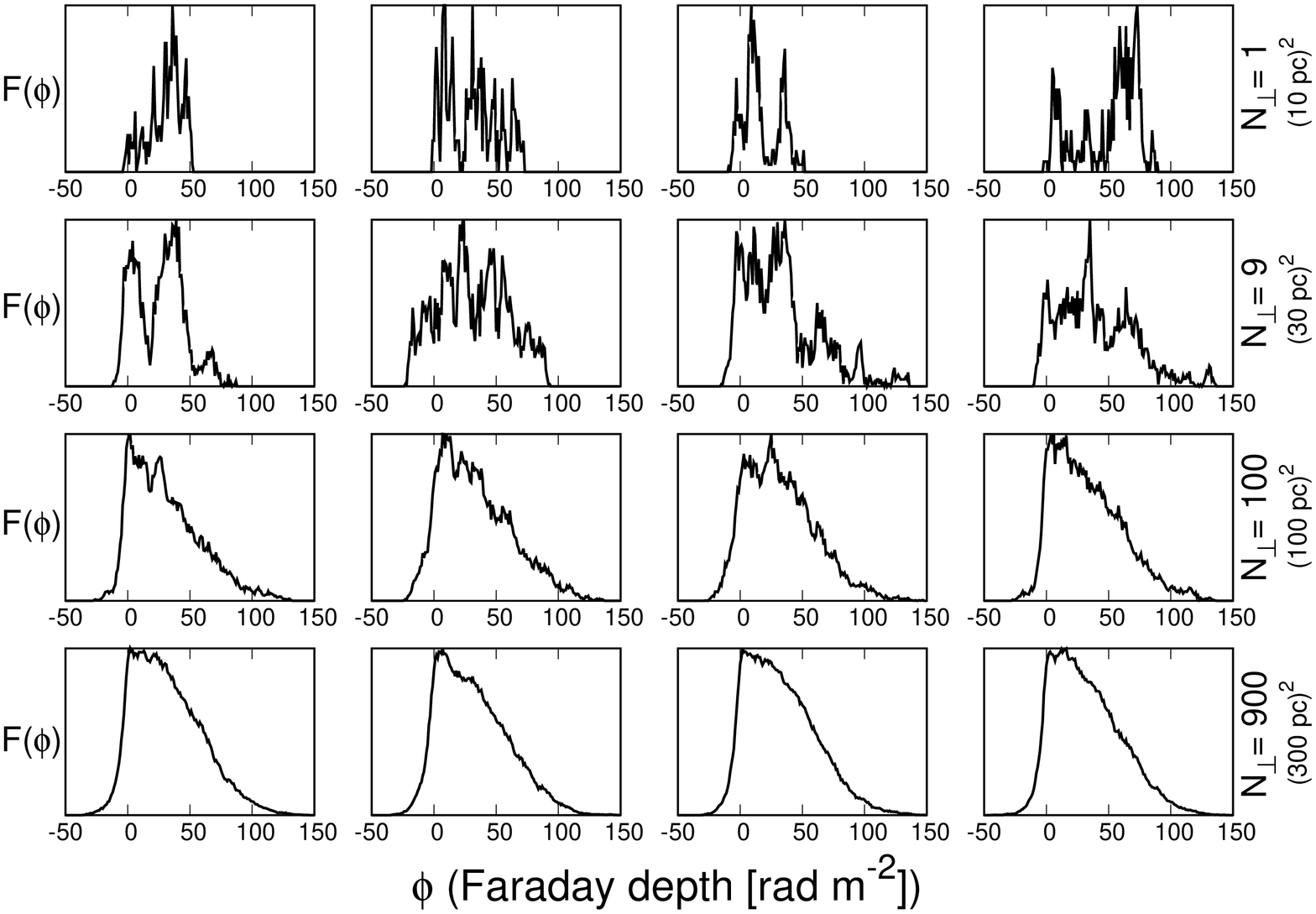}
\end{center}
\vskip 0.6cm
\caption{Same as Figure \ref{fig01}, but with $L_{\rm SH}=0.5$~kpc (left) and 2.0~kpc (right); $B_{\rm coh} = 1~{\rm \mu G}$ and $L_{\rm cell} = 10$ pc. \label{fig06}}
\end{figure*}

Figure \ref{fig06} compares $F(\phi)$ for $L_{\rm SH}=0.5$ and 2.0~kpc; other parameters are $B_{\rm coh} = 1\ {\rm \mu G}$ and $L_{\rm cell} = 10$ pc. The change in $L_{\rm SH}$ affects to the number of cells (or the number of coherence lengths of turbulent magnetic field) along the LOS; $N_\parallel=100$ for $L_{\rm SH}=0.5$~kpc and $N_\parallel=400$ for $L_{\rm SH}=2.0$~kpc. $F(\phi)$ converges to universal shapes for $N_\perp \gtrsim 100$ as before, but the shapes are different for different $L_{\rm SH}$, because $\phi$ spans a larger range with larger $L_{\rm SH}$.

\begin{figure*}
\figurenum{7}
\begin{center}
\vskip -0.4cm
\hskip -0.7cm
\includegraphics[width=0.34\textwidth,angle=-90]{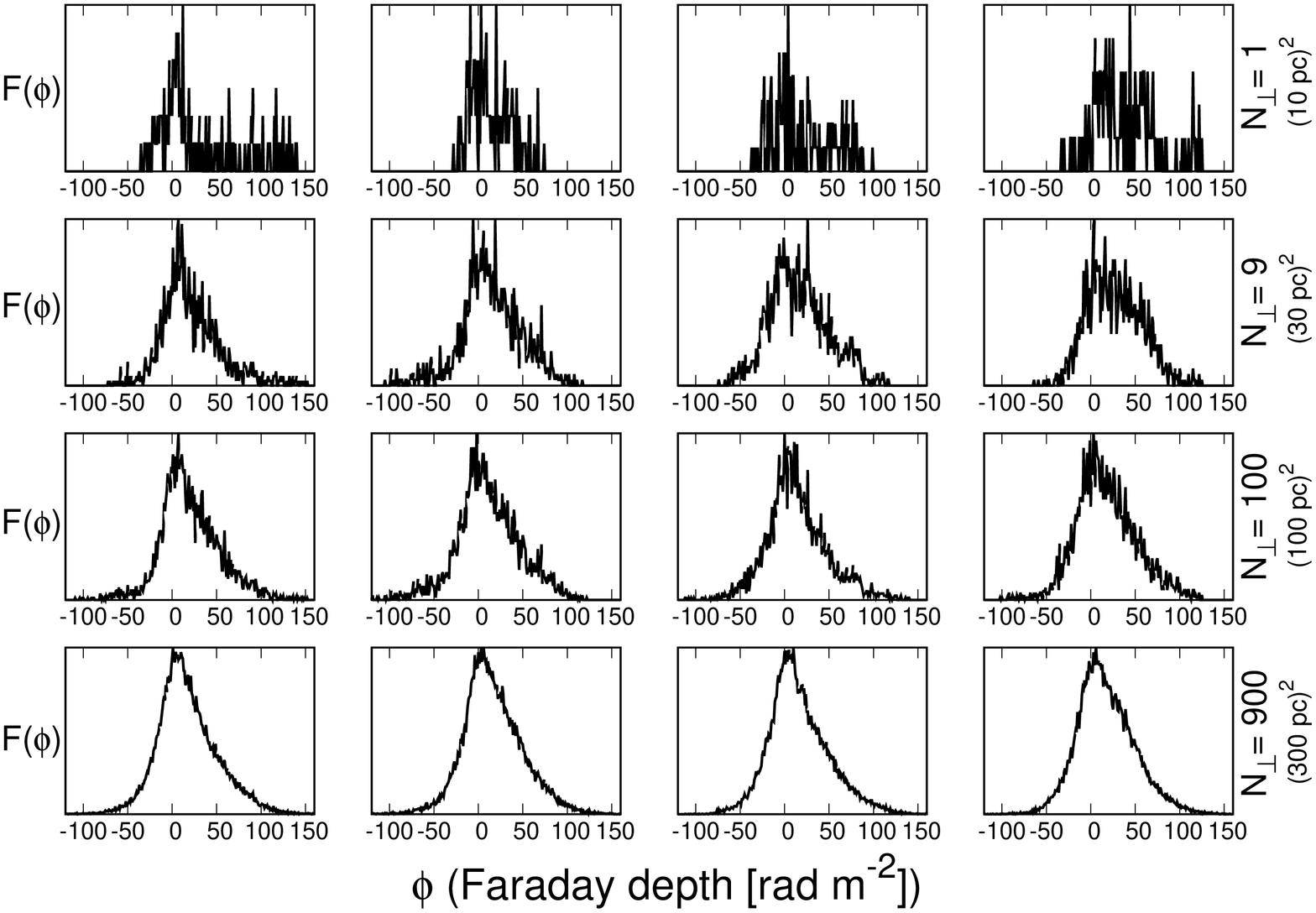}
\hskip 0.6cm
\includegraphics[width=0.34\textwidth,angle=-90]{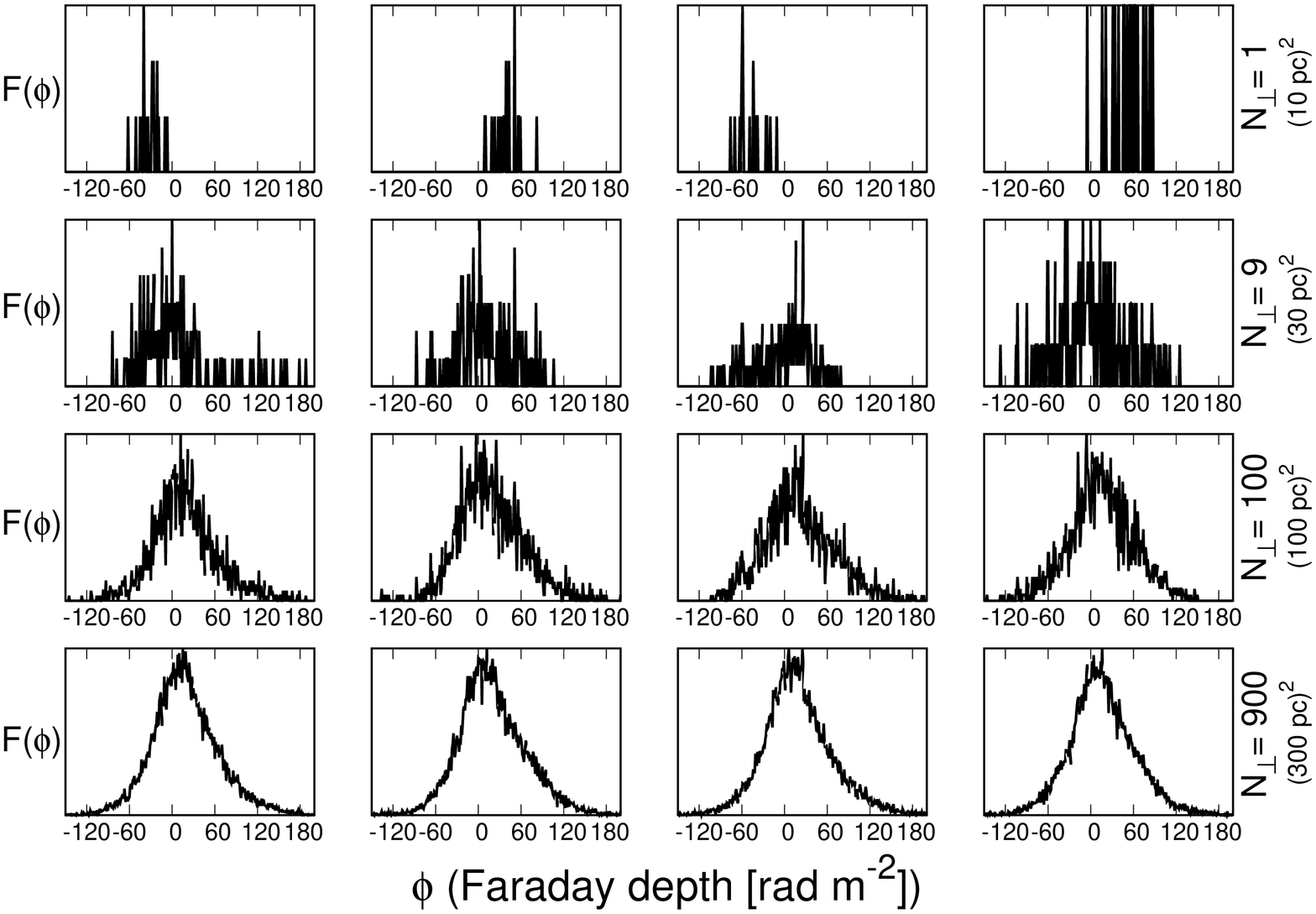}
\end{center}
\vskip 0.6cm
\caption{Same as Figure \ref{fig01}, but with $L_{\rm cell}=50$~pc (left) and 100~pc (right); $B_{\rm coh} = 1~{\rm \mu G}$ and $L_{\rm SH}=1$~kpc. \label{fig07}}
\end{figure*}

Figure \ref{fig07} compares $F(\phi)$ for $L_{\rm cell}=50$ and 100~pc; other parameters are $B_{\rm coh} = 1\ {\rm \mu G}$ and $L_{\rm SH}=1.0~{\rm kpc}$. The number of cells along the LOS is $N_\parallel=40$ for $L_{\rm cell}=50$~pc and $N_\parallel=20$ for $L_{\rm cell}=100$~pc. The figure shows $F(\phi)$ with $N_\perp$ corresponding to the covering area of ${\rm (100~pc)^2}$ ($N_\perp=4$ for $L_{\rm cell}=50~{\rm pc}$ and $N_\perp=1$ for $L_{\rm cell}=100~{\rm pc}$) and ${\rm (300~pc)^2}$ ($N_\perp=36$ for $L_{\rm cell}=50~{\rm pc}$ and $N_\perp=9$ for $L_{\rm cell}=100~{\rm pc}$) to be compared with that for $L_{\rm cell}=10~{\rm pc}$, as well as $F(\phi)$ with $N_\perp=100$ and 900. For convergence, again $N_\perp \gtrsim 100$ is required. But for $L_{\rm cell}=100~{\rm pc}$, even $N_\perp=100$ does not produce smooth $F(\phi)$, because $N_\parallel$ is too small, or the path length does not include enough number of coherence lengths. The converged shape, on the other hand, only weakly depends on $L_{\rm cell}$.

\section{Analytic Faraday spectrum}
\label{sec3}

We next analytically derive $F(\phi)$ and the shape-characterizing parameters in the limit of large $N_\perp$, or large numbers of LOSs, and give interpretations on the results presented in the previous section. The Faraday depth up to the $n$-th layer along the LOS can be written as
\begin{equation}
\phi_n
= \sum_{j=1}^n \left(\Delta \phi_{\rm coh} + \Delta \phi_{\rm rand}^j\right)
= n \Delta\phi_{\rm coh} + \sum_{j=1}^n \Delta \phi_{\rm rand}^j. \label{eq:phi}
\end{equation}
Here, $\Delta\phi_{\rm coh} = K n_e B_{\rm coh} L_{\rm cell}$ is the contribution from the coherent component of $B_{\parallel}$, which is same for all layers; $\Delta\phi_{\rm rand}^j = K n_e B_{\rm rand}^j L_{\rm cell}$ is from the random component of the $j$-th layer. The mean and variance of the random part are
\begin{eqnarray}
&& \left\langle \sum_{j=1}^n \Delta \phi_{\rm rand}^j \right\rangle = 0 \\
&& \left\langle \left( \sum_{j=1}^n \Delta \phi_{\rm rand}^j \right)^2 \right\rangle
   = \left\langle \sum_{j=1}^n \left( \Delta \phi_{\rm rand}^j \right)^2 \right\rangle
   = n \sigma_\phi^2,\ \label{eq:phi_rand_var}
\end{eqnarray}
where $\sigma_\phi^2 = K^2 n_e^2 \sigma_B^2 L_{\rm cell}^2$ with $\langle B_{\rm rand}^2 \rangle = \sigma_B^2$. We assume that there is no correlation between $B_{\rm rand}$ of different layers.

\begin{figure*}
\figurenum{8}
\begin{center}
\vskip -1.3cm
\hskip -1.7cm
\includegraphics[width=0.52\textwidth,angle=-90]{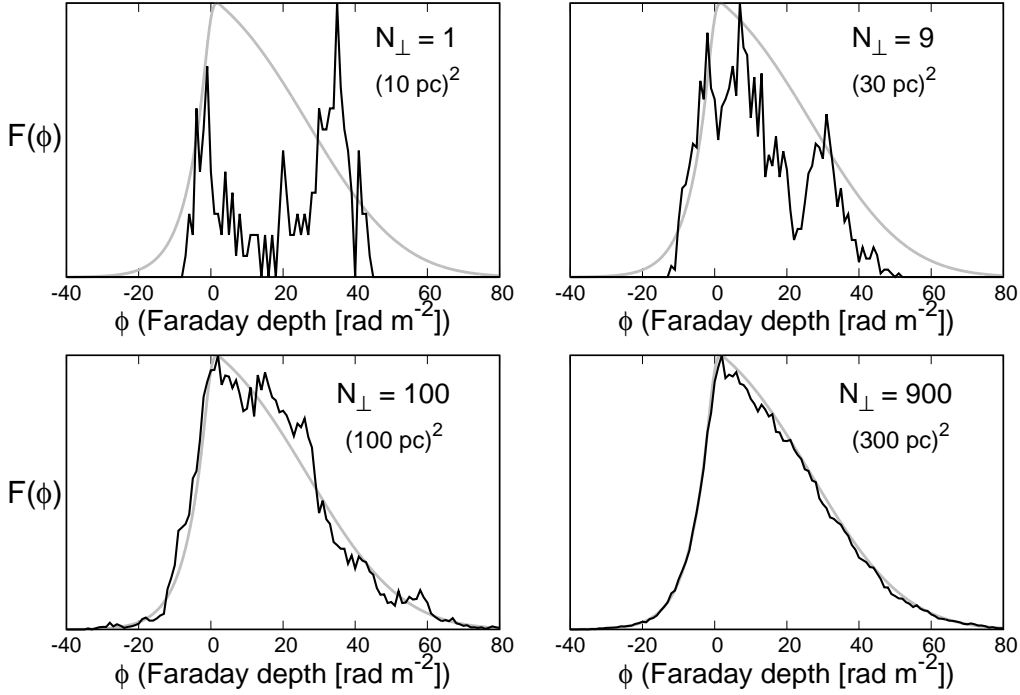}
\end{center}
\vskip 0.9cm
\caption{{\color{red}Simulated Faraday spectrum, $F(\phi)$, as a function of Faraday depth, $\phi$,} for $N_\perp=$ 1, 9, 100, and 900 (black lines), overlaid with the analytically reproduced spectrum (gray lines); $B_{\rm coh} = 1~{\rm \mu G}$, $L_{\rm cell} = 10$ pc, and $L_{\rm SH} = 1$ kpc. The simulated spectra are the same as those in Figure \ref{fig01}, but different realizations. \label{fig08}}
\end{figure*}

We further assume that the polarized synchrotron emissivity and the polarization angle are uniform throughout the computational domain (see section \ref{sec2.1}). Then, the $j$-th layer's contribution to Faraday spectrum, $P_j(\phi)$, is proportional to the probability distribution of the Faraday depth of the $j$-th layer, and $F(\phi)$, aside from the overall normalization, is given by
\begin{equation}\label{eq:sumPana}
F(\phi) \propto \sum_{j=1}^{N_\parallel} P_j(\phi).
\end{equation}
The functional form of $P_j(\phi)$ reflects the characteristics of the probability distribution. In the limit of large $N_\perp$, the central limit theorem dictates that $P_j(\phi)$ approaches to the normal distribution with $j \Delta \phi_{\rm coh}$ as the mean and $j \sigma_\phi^2$ as the variance;
\begin{equation}\label{eq:Pana}
P_j(\phi)
= \frac{1}{\sqrt{2\pi j}\sigma_\phi}\exp{\left[ -\frac{(\phi-j\Delta\phi_{\rm coh})^2}{2j\sigma_\phi^2} \right]}.
\end{equation}
That is, the Faraday spectrum is approximated to a sum of many Gaussian functions with different means and variances. Figure \ref{fig08} shows comparisons of simulated $F(\phi)$ with the spectrum in Equation (\ref{eq:sumPana}) and (\ref{eq:Pana}) for $B_{\rm coh} = 1~{\rm \mu G}$. As $N_\perp$ increases, the statistical fluctuations due to the turbulence magnetic field reduce and simulated $F(\phi)$ approaches to the analytical solution.

\begin{figure*}
\figurenum{9}
\begin{center}
\vskip -1.2cm
%\hskip -1.2cm
\includegraphics[width=0.34\textwidth,angle=-90]{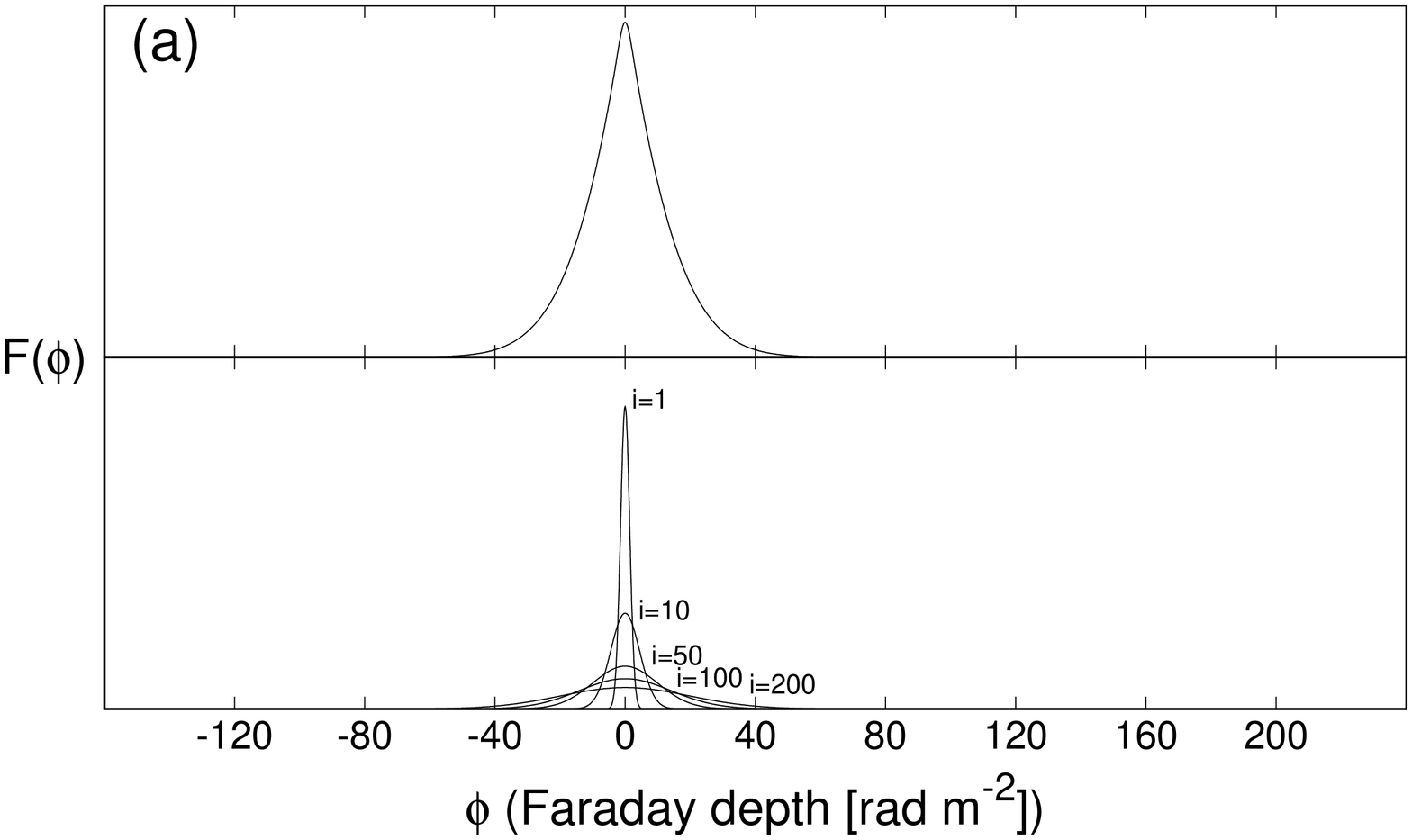}
\hskip -0.3cm
\includegraphics[width=0.34\textwidth,angle=-90]{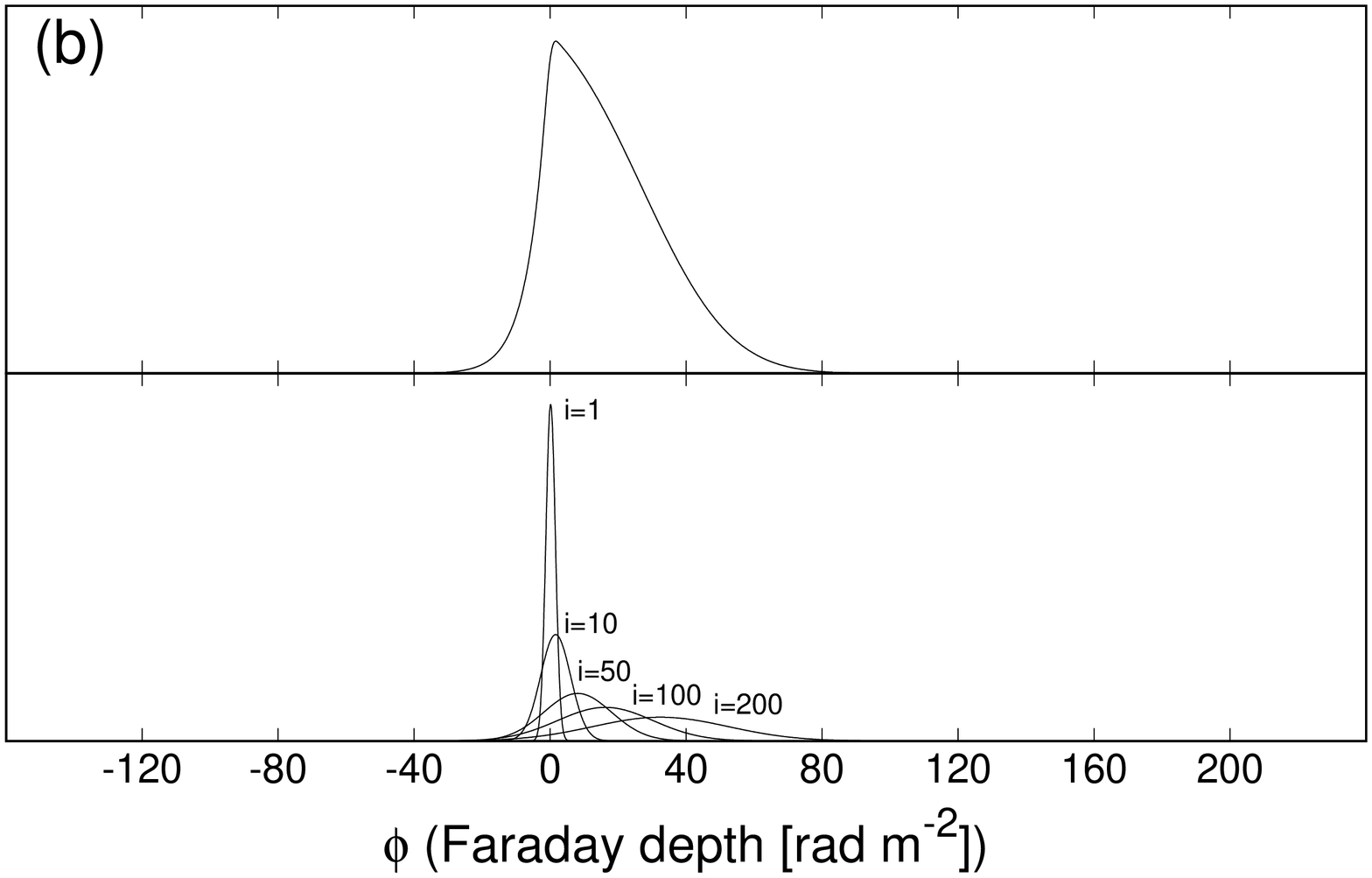}
%\end{center}
%\begin{center}
%\hskip -1.2cm
\includegraphics[width=0.34\textwidth,angle=-90]{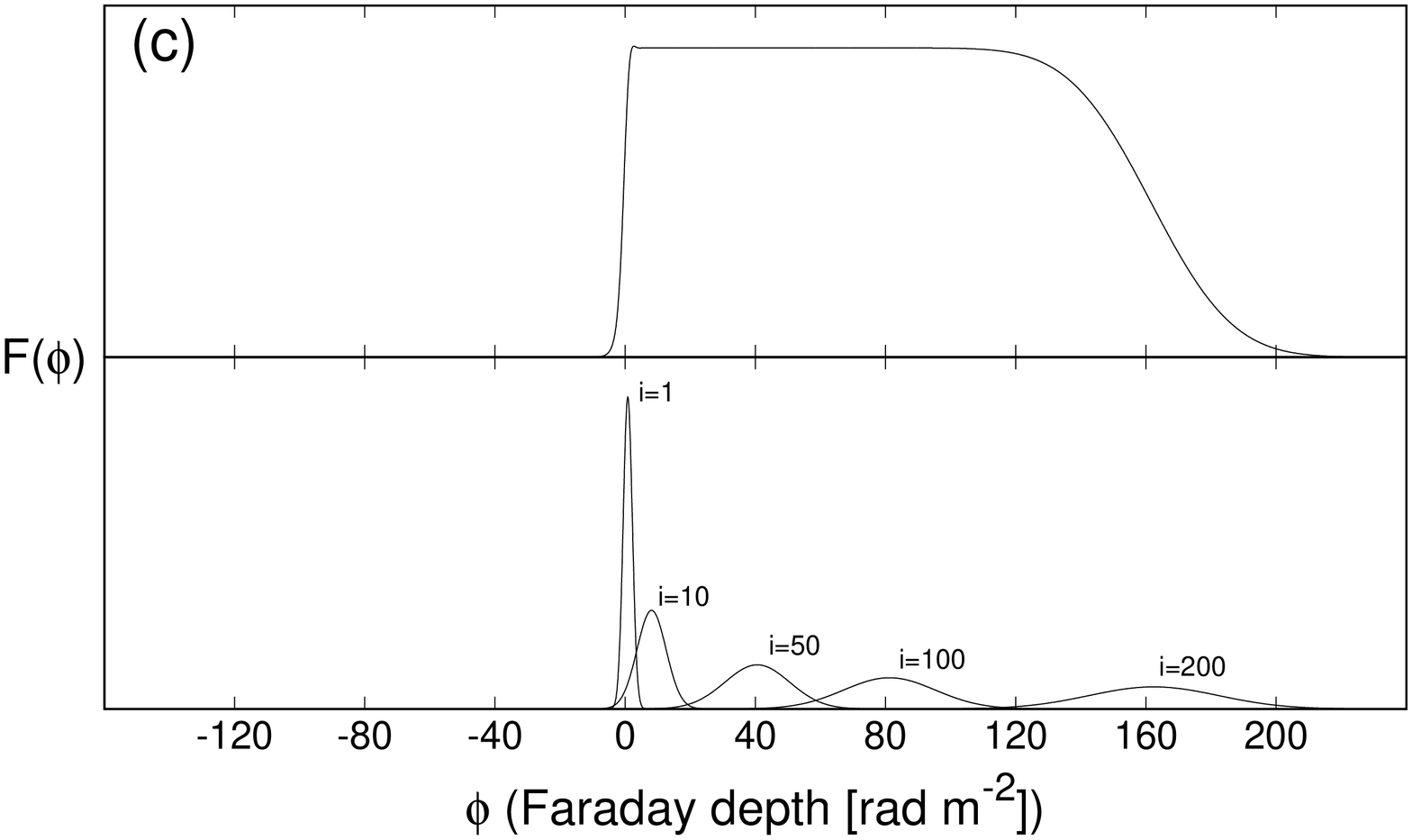}
\hskip -0.2cm
\includegraphics[width=0.34\textwidth,angle=-90]{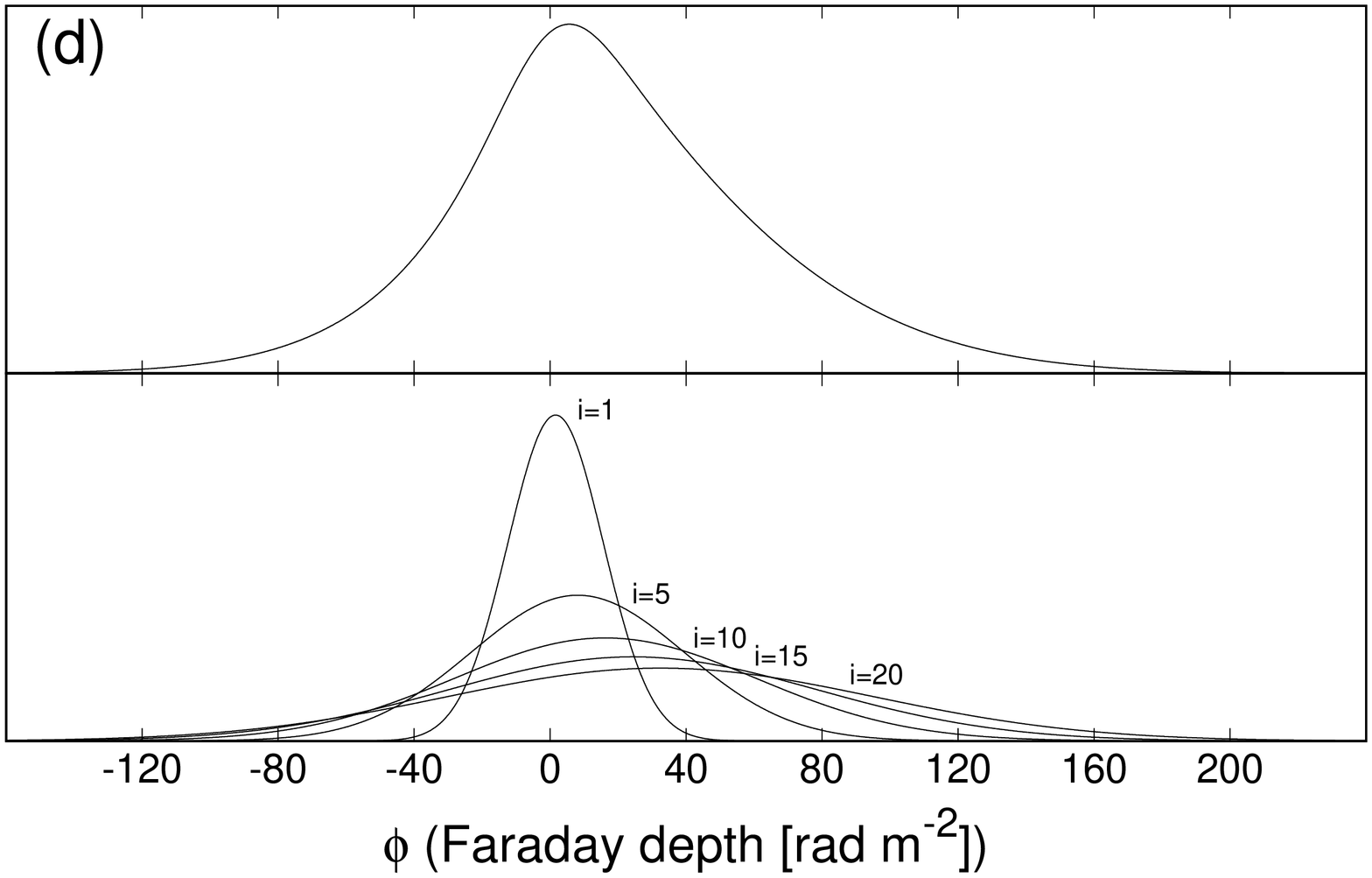}
\end{center}
\vskip 0.3cm
\caption{Analytical Faraday spectra with $(L_{\rm cell}[{\rm pc}],B_{\rm coh}[\mu{\rm G}])=$ $(10,0)$, $(10,1)$, $(10,5)$ and $(100,1)$, shown in (a), (b), (c) and (d), respectively; $L_{\rm SH}=1$~kpc. The lower parts of panels show contributions from a number of different layers using Equation (\ref{eq:Pana}). The spectra in the upper parts are the sums of contributions along the LOS.\label{fig09}}
\end{figure*}

Figure \ref{fig09} illustrates how the specific shape of $F(\phi)$ is induced for different parameters of $(L_{\rm cell}[{\rm pc}],B_{\rm coh}[\mu{\rm G}])=$ $(10,0)$, $(10,1)$, $(10,5)$ and $(100,1)$ in (a), (b), (c) and (d), respectively. When $B_{\rm coh} = 0~{\rm \mu G}$, the contribution from each layer is the Gaussian with zero mean, but the variance increases with increasing $j$. As a consequence, $F(\phi)$ becomes symmetric about $\phi = 0$ with zero skewness and leptokurtic with positive kurtosis (see also Figure \ref{fig03}). With non-zero $B_{\rm coh}$, the mean of the Gaussian also increases as $j$ increases. So $F(\phi)$ becomes skewed toward positive $\phi$, and the shape changes from leptokurtic to platykurtic as $B_{\rm coh}$ increases, as shown in Figure \ref{fig09} (b) and (c) (also in Figure \ref{fig03}). In the figures, $F(\phi)$ for large positive $\phi$ represents emissions from the far side of the computational box. Their contribution for large $\phi$ is small, because emissions from far side experience Faraday rotation due to the turbulent fields in nearer layers and spread in the $\phi$ space. We note that the polarization angle of emissions is assumed to be uniform in our model, and any depolarizations are not included in our calculation (see section \ref{sec2.1}). For $B_{\rm coh}$ comparable to or larger than $\sigma_B$, $F(\phi)$ stretches over a large range of $\phi$, and the skewness decreases. For larger $L_{\rm cell}$, shown in Figure \ref{fig09} (d), the variance from each layer is larger, but the number of layers is smaller. So the shape becomes relatively more symmetric.

Once the Faraday spectrum is given as in Equations (\ref{eq:sumPana}) and (\ref{eq:Pana}), the width, skewness and kurtosis can be analytically calculated as (see Appendix),
\begin{eqnarray}
\sigma^2
&=& \frac{N_\parallel^2-1}{12}\Delta\phi_{\rm coh}^2+\frac{N_\parallel+1}{2}\sigma_\phi^2 \nonumber \\
&\rightarrow& \frac{N_\parallel^2}{12} \Delta \phi_{\rm coh}^2
              + \frac{N_\parallel}{2} \sigma_\phi^2 \nonumber \\
&=& \frac{K^2 n_e^2 N_\parallel L_{\rm cell}^2}{12}
  \left( N_\parallel B_{\rm coh}^2 + 6 \sigma_B^2 \right) \label{eq:anasigma}, \\
\gamma_{\rm s}
&=& \frac{N_\parallel(N_\parallel^2-1) \Delta \phi_{\rm coh} \sigma_\phi^2}{4\left[ \frac{N_\parallel^2-1}{12}\Delta\phi_{\rm coh}^2+\frac{N_\parallel+1}{2}\sigma_\phi^2 \right]^{3/2}} \nonumber \\
&\rightarrow& \frac{6 \sqrt{3 N_\parallel} B_{\rm coh} \sigma_B^2}{(N_\parallel B_{\rm coh}^2 + 6 \sigma_B^2)^{3/2}} \nonumber \\
&=& {\rm sign}(B_{\rm coh}) \frac{6 \sqrt{3} \sqrt{\alpha}}{(\alpha + 6)^{3/2}} \label{eq:anaskew}, \\
\gamma_{\rm k}
&=& \frac{-\frac{N_\parallel^4-1}{120}\Delta\phi_{\rm coh}^4+\frac{N_\parallel^2-1}{4}\sigma_\phi^4}{\left[ \frac{N_\parallel^2-1}{12}\Delta\phi_{\rm coh}^2+\frac{N_\parallel+1}{2}\sigma_\phi^2 \right]^2} \nonumber \\
&\rightarrow& -\frac{6}{5}\frac{N_\parallel^2B_{\rm coh}^4-30\sigma_B^4}{(N_\parallel B_{\rm coh}^2+6\sigma_B^2)^2}
= -\frac{6}{5} \frac{\alpha^2 - 30}{(\alpha + 6)^2},
 \label{eq:anakurt}
\end{eqnarray}
where ``$\rightarrow$" denotes the limit of $N_\parallel \gg 1$ ($N_\parallel = 10 - 400$ in our model; see section \ref{sec2}) and $\alpha = N_\parallel B_{\rm coh}^2/\sigma_B^2$. The black lines of Figure \ref{fig04} and \ref{fig05} show these analytical solutions, which well reproduce simulated results.

We can learn the followings.
(i) The width increases with increasing $B_{\rm coh}$, $\sigma_B$, and the coherence length $L_{\rm cell}$.
(ii) Both the skewness and kurtosis are expressed with a single parameter, $\alpha = N_\parallel B_{\rm coh}^2/\sigma_B^2$, which represents the relative importance of coherent to random fields (see Equations (\ref{eq:phi}) and (\ref{eq:phi_rand_var})).
(iii) The skewness is zero for $\alpha = 0$ and also for $\alpha \rightarrow \infty$, and its sign is determined by the sign of $B_{\rm coh}$.
(iv) The kurtosis changes from +1 (leptokurtic) for $\alpha = 0$ to -6/5 (platykurtic) for $\alpha \rightarrow \infty$.
Hence, if the width, skewness, and kurtosis of $F(\phi)$ are obtained from observations, we may be able to get the information such as the strengths of the global and random components of the magnetic field parallel to LOSs as well as the coherence length of the turbulent magnetic field (see section \ref{sec5}). We point that the shape-characterizing parameters of ``intrinsic'' $F(\phi)$ are expressed with $N_\parallel$, $B_{\rm coh}$, and $\sigma_B$ as shown in Equations (\ref{eq:anasigma}) - (\ref{eq:anakurt}), and do not depend on the observation frequency coverage.

\section{Faraday spectrum for turbulent magnetic field with power-law energy spectrum}
\label{sec4}

\begin{figure*}
\figurenum{10}
\begin{center}
\vskip -0.2cm
\hskip -1.6cm
\includegraphics[width=0.52\textwidth,angle=-90]{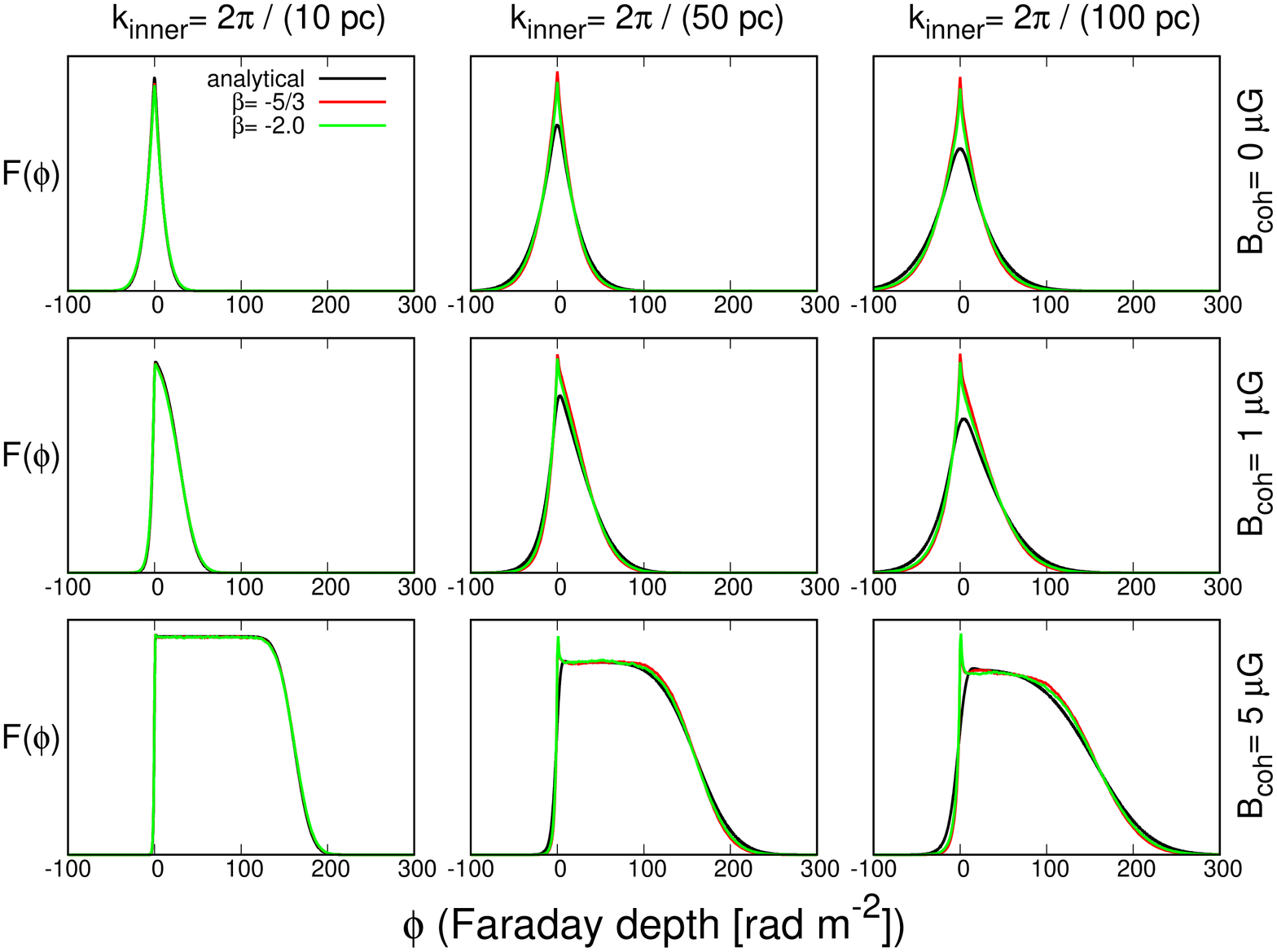}
\end{center}
\vskip 1.3cm
\caption{{\color{red}Simulated Faraday spectrum, $F(\phi)$, as a function of Faraday depth, $\phi$,} for the turbulent magnetic field reproduced with power-law spectra; $B_{\rm coh}=0,1,5~\mu{\rm G}$ from top to bottom, and $\sigma_B = 15/\sqrt{3}\ \mu$G. The spectra with different outer scales, $L_{\rm outer}$, and power-law slopes, $\beta$, are shown. The black lines are the analytic spectra of section \ref{sec3} for $\beta=-5/3$. See the main text for further details.\label{fig10}}
\end{figure*}

We also consider a bit more realistic magnetic field model, where turbulent magnetic field is represented by the energy spectrum of two power-laws, such as 
\begin{eqnarray}
\left\{
\begin{array}{cc}
E_B(k) \propto k^\alpha & ~(k\leq k_{\rm innner}) \\
E_B(k) \propto k^\beta   & ~(k> k_{\rm inner}).
\end{array}  \right.
\end{eqnarray}
The outer scale in real space (that corresponds to an inner scale in Fourier space), $L_{\rm outer} \equiv 2\pi / k_{\rm inner}$, is set to be $10 - 100$ pc (see section \ref{sec2.1}). The slope for $k\leq k_{\rm inner}$, $\alpha$, is fixed as 2 \citep[see, e.g.,][]{lesi97}, while for $k> k_{\rm inner}$ a range of values, $-2 \leq \beta \leq -1.5$, are considered. $\beta=-5/3$ is the Kolmogorov slope, which is close, for instance, to the power spectrum slope of the interstellar electron density \citep[see, e.g.,][]{arms81}. 

In a box of size $L_{\rm box} = 2$ kpc, divided into (512)$^3$ grid zones (so the grid size is $\sim$ 4~pc), a 3D turbulent magnetic field is constructed, as follows. The Fourier components, satisfying ${\vec k}\cdot{\vec B_k}$ (so ensuring ${\vec\nabla}\cdot {\vec B}=0$ in the real space), are drawn from a Gaussian random field in the Fourier space. Their relative amplitude is determined by the above spectrum. The components are converted to quantities in the real space by Fourier transform and then added. The absolute amplitude is tuned in such a way that the resulting 3D magnetic field has the rms value of 15 $\mu$G. Then, the LOS component is taken as $B_{\rm rand}$, which has $\sigma_B = 15 / \sqrt{3}$ $\mu$G. The thermal electron density, $n_e$, synchrotron emissivity, and polarization angle are assumed to be uniform within the computational domain, as in Section \ref{sec2}.

Figure \ref{fig10} shows simulated $F(\phi)$ with $B_{\rm coh} =$ 0, 1, and 5 $\mu$G from top to bottom, for different $L_{\rm outer}$ and $\beta$. The profiles of $F(\phi)$ are smooth, since $F(\phi)$ is obtained with $N_\perp = 512^2$ covering $(2\ {\rm kpc})^2$ region. Although not shown here, once the covering region is sufficient large, specifically larger than $\sim (10 L_{\rm int})^2$ (see below for the definition of $L_{\rm int}$), $F(\phi)$ converges, similarly as discussed in section \ref{sec2}. The shape of $F(\phi)$ changes sensitively by changing $B_{\rm coh}$, or more precisely its strength relative to $\sigma_B$, as well as by changing $L_{\rm outer}$. On the other hand, the dependence on $\beta$ is weak in the range of $\beta$ considered.

The black lines of Figure \ref{fig10} show analytically constructed $F(\phi)$ of section \ref{sec3} with the integral scale length,
\begin{equation}
L_{\rm int} = 2\pi \int \frac{E_B(k)}{k} dk \bigg / \int E_B(k) dk,
\end{equation}
for $\beta=-5/3$, as the coherence length (that is, $L_{\rm int}$ used for $L_{\rm cell}$ in Equations (\ref{eq:sumPana}) and (\ref{eq:Pana})), and correspondingly, with $N_\parallel = L_{\rm box} / L_{\rm int}$. Note that, $L_{\rm int} = 0.5 - 0.75$ $L_{\rm outer}$ for $-2 \leq \beta \leq -1.5$. The analytically constructed spectra well fit to simulated ones. This is expected, since it is known that the variance of RM can be expressed with $L_{\rm int}$ for the coherence length of turbulent magnetic field \citep[see, e.g.,][]{cho09}. This result implies that even if the turbulent part of the galactic magnetic field is described by power-law spectra, once a smooth profile of $F(\phi)$ is obtained through observations of multiple LOSs, the width, skewness and kurtosis may be used to retrieve the strength of the global and random components of $B_{\parallel}$ as well as the integral scale length of the turbulent magnetic field.

\section{Summary and discussion}
\label{sec5}

The study of cosmic magnetic field using Faraday tomography involves not only the reconstruction of Faraday spectrum, $F(\phi)$, through observation of polarization spectrum, but also the extraction of magnetic field information from $F(\phi)$. The latter part, however, often turns out to be complicated, mainly because of the turbulent component of magnetic field; it causes the relation between the Faraday depth and the physical depth to be non-trivial and produces Faraday forest \citep{fri11,bec12}, many small-scale features, in $F(\phi)$. Our previous work \citep{ide14b} showed that $F(\phi)$ calculated with a realistic model for the Milky Way \citep{aka13} has Faraday forest superposed on large-scale diffuse emission. We also found that $F(\phi)$ can have significantly different shapes for different configurations of turbulence, despite the global parameters of the model are fixed. But in \citet{ide14b}, the interpretation of $F(\phi)$ was limited, due to its complicated behavior. In this work, we studied $F(\phi)$ of face-on spiral galaxies with the magnetic fields described with simpler, toy models, and tried to numerically as well as analytically interpret $F(\phi)$. We investigated how $F(\phi)$ along multiple LOSs, covering a small region where the properties of magnetic field and other quantities such as thermal and CR electron densities are assumed to be uniform, can be used in Faraday tomography study.

With the turbulent magnetic field described as a random field with single coherence length, we numerically showed that small-scale features in $F(\phi)$ are smoothed out and the shape of $F(\phi)$ converges, if $F(\phi)$ is obtained with LOSs covering a region of $\gtrsim (10\ {\rm coherence\ length)^2}$ in the sky. Note that this explains why we failed to get converged $F(\phi)$ in \citet{ide14b}; with $L_{\rm int} \sim$ 75 pc, the covering region of $({\rm 500~pc})^2$ is smaller than the requirement for convergence. Also note that we do not need very high angular resolutions of {\color{red}radio} interferometers to apply this method, in the sense that the observed field should be much larger than the coherence length of turbulent field to smooth out the small-scale features in $F(\phi)$.

We then analytically showed that the converged $F(\phi)$ can be expressed as a sum of Gaussian functions with $j\Delta\phi_{\rm coh}$ as the mean and  $j\sigma_\phi^2$ as the variance along LOSs; $j\Delta\phi_{\rm coh}$ is the RM up to the $j$-th layer due to the coherent component of $B_{\parallel}$, $B_{\rm coh}$, and $j\sigma_\phi^2$ is the variance of RM due the random component of $B_{\parallel}$, $B_{\rm rand}$. The analytical expression was derived using the central limit theorem. Then, the shape-characterizing parameters, that is, the width, skewness, and kurtosis of $F(\phi)$ are given as simple functions of the strength of $B_{\rm coh}$ and the variance and coherence length of $B_{\rm rand}$.

With the turbulent magnetic field reproduced with power-law spectra, the same results are obtained, once the coherence length is replaced with the integral length of the turbulent magnetic field.

Our results suggest a way to extract quantities such as the strength and coherence length of the vertical magnetic field in face-on spiral galaxies with Faraday tomography. We point that $F(\phi)$ along single LOS and $F(\phi)$ constructed with multiple LOSs can be used differently. While $F(\phi)$ along single LOS can tell us, for instance, the existence of turbulent field, $F(\phi)$ along multiple LOSs can provide us with the global properties of magnetic field such as the strength and coherence length.

Our analytic expressions could be applied to interpret the results of other works. For instance, \citet{fri11} calculated $F(\phi)$ including both regular and turbulent fields, and got small skewness. They assumed the Gaussian distribution of large-scale field with the peak strength of $\sim$ 2.0 $\mu$G, and the rms value of small-scale turbulent field with Kolmogorov spectrum is twice that of large-scale field. If we estimate $N_\parallel=200$, $B_{\rm coh}=2.0\ {\rm \mu G}$ and $\sigma_B^2=16/3\ {\rm(\mu G)^2}$ (so the rms strength of random field is 4 $\mu$G) for simplicity, $\alpha \sim$ 150. From Equation (\ref{eq:anaskew}), note that the skewness is large only for $\alpha$ around unity, that is, only when the contributions of coherent and turbulent fields to $\phi$ are comparable. The models adopted in \cite{fri11} result in small skewness, i.e., $\gamma_s \sim$ 0.065 for $\alpha \sim$ 150, mainly because the contribution of coherent field is much larger than that of turbulent field.

The $\alpha$ parameter is composed of three quantities, $N_\parallel$, $B_{\rm coh}$ and $\sigma_B$. While it would be useful if we could separate them from observables such as skewness and kurtosis, that should not be easy mainly because the quantities degenerate. For instance, any combinations of three quantities providing the same $\alpha$ value result in the same skewness and kurtosis. However, the width of $F(\phi)$ is large if $N_\parallel$ and $B_{\rm coh}$ are large, regardless of $\sigma_B$ value. Hence, we may be able to understand how the three parameters depend on the shape-characterizing parameters. We will leave the exploration of this as a future work.

In this work, we ignored possible differences between disk and halo (or thick disk). Observations suggested that the halo magnetic field would have a topology very different from that of disk \citep[e.g.,][]{fle11}. If the component of the halo magnetic field parallel to the LOS is mostly turbulent, such field may lead to Faraday dispersion, which broadens and weakens the signals seen in $F(\phi)$, and $F(\phi)$ would become further complicated. If the component is mostly coherent and halo does not contribute to polarized emission, $F(\phi)$ only shifts in the $\phi$ space. The impact of halo to $F(\phi)$ will depend on the amount of polarized emission. If the halo emission is as large as that of disk, the observed spectrum may suffer substantial wavelength-independent depolarization, since the perpendicular components of halo and disk fields would be in general not aligned with each other. However, observations showed that the distribution of radio emission from halos of edge-on spiral galaxies can be described by exponential function, for instance, with the scale heights of {\color{red}about} 1.8~kpc \citep{kra09}. This suggest that the halo emission {\color{red}is} small compared to that of disk.

Finally, we consider the work presented here to be the first step toward understanding the intrinsic characteristics of $F(\phi)$, and thus it needs to be further sophisticated with more realistic treatments of galactic magnetic field. In addition, when $F(\phi)$ is constructed from an observed polarization spectrum, the effects such as false signal in RM CLEAN \citep{far11,kum14,miy16} as well as the limited frequency coverage and noises in observation need to be considered. For instance, the shape of $F(\phi)$ could depend on wavelength because of imperfect Fourier transform due to the limited sampling of squared-wavelength. Also, the resolution in Faraday depth space, which is determined by the $\lambda^2$ coverage \citep{bd05}, becomes important for the method presented here to be applied. In the case of large $B_{\rm coh}$ like 5 $\mu$G (e.g. Figure \ref{fig09} (c)), the resolution of $\lesssim {\rm a~few~10~rad~m^{-2}}$ may be enough to calculate the shape-characterizing parameters. Full ASKAP (700 - 1800 MHz), giving a $\sim 22~{\rm rad~m^{-2}}$ resolution, would then be good enough. On the other hand, when $B_{\rm coh}$ is smaller like 1 $\mu$G (e.g. Figure \ref{fig09} (b)), the resolution of $\lesssim {\rm 10~rad~m^{-2}}$ seems to be necessary. Upgraded GMRT (e.g. 300 - 900 MHz), which gives a $\sim 4~{\rm rad~m^{-2}}$ resolution, could then be used. Furthermore, if we try to apply the method to galaxies with much weaker fields such as the Milky Way, where {\color{red}the vertical $B_{\rm coh}$ at the solar radius} is up to $\sim 0.3~\mu$G \citep{tay09,mao10} and the random field is $\sim 5 \ \mu$G \citep{orl13} toward the direction of the Galactic poles, we need a much higher resolution due to the smaller width of $F(\phi)$. Hence, LOFAR (e.g., 120 - 240 MHz, High Frequency Band), giving a $\lesssim 1~{\rm rad~m^{-2}}$ resolution would be necessary. Indeed, LOFAR so far has not detected extended polarized emissions from spiral galaxies at frequencies below 200 MHz, probably because of Faraday depolarization. {\color{red}We may have to wait for SKA.} Thus, it is necessary to examine how well shape-characterizing parameters will be determined after considering observational effects. We will leave these as future works.

\acknowledgments
We thank the anonymous referee for constructive comments. S.I. was supported by the National Research Foundation of Korea through grant 2007-0093860. T.A. was supported by JSPS KAKENHI Grant Numbers 15K17614 and 15H03639. K.T. was supported by Grants-in-Aid from the Ministry of Education, Culture, Sports, Science, and Technology (MEXT) of Japan, Nos. 24340048, 26610048, 15H05896, and 16H05999. D.R. was supported by the National Research Foundation of Korea through grant 2016R1A5A1013277.

\appendix

\section{Calculation of shape-characterizing parameters}

For the derivation of the width, skewness and kurtosis in Equations (\ref{eq:anasigma}) $-$ (\ref{eq:anakurt}), we employ $F(\phi)$ in Equations (\ref{eq:sumPana}) and (\ref{eq:Pana}) and replace the summation in Equations (\ref{eq:sigma}) $-$ (\ref{eq:kurt}) with the integration. That is,
\begin{eqnarray}
\sum_l |F(\phi_l)|
&\rightarrow& \int_{-\infty}^{+\infty}|F(\phi)| d\phi \nonumber \\
&=& \int_{-\infty}^{+\infty} \sum_{j=1}^{N_\parallel} \frac{1}{\sqrt{2\pi j}\sigma_\phi}\exp{\left[ -\frac{(\phi-j\Delta\phi_{\rm coh})^2}{2j\sigma_\phi^2} \right]} d\phi \nonumber \\
&=& \sum_{j=1}^{N_\parallel} \int_{-\infty}^{+\infty} \frac{1}{\sqrt{2\pi j}\sigma_\phi}\exp{\left[ -\frac{(\phi-j\Delta\phi_{\rm coh})^2}{2j\sigma_\phi^2} \right]} d\phi \nonumber \\
&=& N_\parallel,
\end{eqnarray}
and
\begin{eqnarray}
\sum_l |F(\phi_l)| \phi_l
&\rightarrow& \int_{-\infty}^{+\infty}|F(\phi)| \phi\ d\phi \nonumber \\
&=& \sum_{j=1}^{N_\parallel} \int_{-\infty}^{+\infty} \frac{1}{\sqrt{2\pi j}\sigma_\phi}\exp{\left[ -\frac{(\phi-j\Delta\phi_{\rm coh})^2}{2j\sigma_\phi^2} \right]} \phi\ d\phi \nonumber \\
&=& \frac{N_\parallel(N_\parallel+1)}{2}\Delta\phi_{\rm coh}.
\end{eqnarray}
Then, the spectrum-weighted average of Faraday depth becomes
\begin{equation}
\mu = \frac{\sum_j |F(\phi_j)| \phi_j}{\sum_j |F(\phi_j)|} = \frac{N_\parallel+1}{2}\Delta\phi_{\rm coh}.
\end{equation}
In the same manner, 
\begin{eqnarray}
\sum_l |F(\phi_l)| (\phi_l-\mu)^2
&\rightarrow& \int_{-\infty}^{+\infty}|F(\phi)| (\phi-\mu)^2 d\phi \nonumber \\
&=& \sum_{j=1}^{N_\parallel} \int_{-\infty}^{+\infty} \frac{1}{\sqrt{2\pi j}\sigma_\phi}\exp{\left[ -\frac{(\phi-j\Delta\phi_{\rm coh})^2}{2j\sigma_\phi^2} \right]} (\phi-\mu)^2 d\phi \nonumber \\
&=& \frac{N_\parallel(N_\parallel+1)(2N_\parallel+1)}{6}\Delta\phi_{\rm coh}^2+N_\parallel(N_\parallel+1)(\sigma_\phi^2-\mu\Delta\phi_{\rm coh})+N_\parallel\mu^2.
\end{eqnarray}
So the width becomes
\begin{equation}
\sigma^2 = \frac{\sum_j |F(\phi_j)| (\phi_j-\mu)^2}{\sum_j |F(\phi_j)|}
= \frac{N_\parallel^2-1}{12}\Delta\phi_{\rm coh}^2+\frac{(N_\parallel+1)\sigma_\phi^2}{2}.
\end{equation}
Similarly, the skewness and kurtosis can be derived.

\end{document}